\documentclass[preprint,12pt]{elsarticle}

\usepackage{amssymb}

\def\Li2#1{{\mathrm{Li}}_2\left(#1\right)}

\begin{document}

\begin{frontmatter}

\title{Off-mass-shell muon anomalous magnetic moment}


\author[bltp,mupoch]{A.B. Arbuzov}
\author[mupoch]{T.V. Kopylova}

\address[bltp]{Bogoliubov Laboratory for Theoretical Physics, JINR, Dubna, 141980 Russia}
\address[mupoch]{Department of Higher Mathematics, University of Dubna, 141980 Dubna, Russia}

\begin{abstract}
Interaction of charged leptons with photons is considered 
for the case when one of the lepton legs is off the mass shell. 
The effect due to off-mass-shell shift in the anomalous magnetic moment 
is computed within one-loop approximation. 
Possible contributions of this effect in the muon $g-2$ measurements
are discussed. 
\end{abstract}


\end{frontmatter}

\section{Introduction}

The anomalous magnetic moment of a charged fermion 
$a_f\equiv(g_f-2)/2$
is defined via its gyromagnetic ratio $g_f$. The Dirac equation gives
$g_f=2$, but quantum corrections can shift it. 
For the electron case, we have the perfect agreement of extremely precise 
experimental~\cite{Beringer:1900zz} results and the corresponding theoretical 
calculations~\cite{Aoyama:2012wj}. That is in fact
a triumph of perturbative quantum electrodynamics. 

On the other hand, the difference between experimental data
and theoretical prediction for the muon anomalous magnetic moment
withstands to our efforts to remove it for many years. 
Very accurate measurements with muons and anti-muons at 
the E821 experiment at BNL~\cite{Bennett:2006fi} gave
\begin{eqnarray}
a^{\mathrm{exp}}_\mu= 11\, 659\, 208.9(5.4)(3.3)\cdot 10^{-10}.
\end{eqnarray}
Various theoretical predictions received within the Standard Model (SM)
are based on well established calculations in the QED~\cite{Aoyama:2012wk} and (electro)weak
sectors~\cite{Czarnecki:2002nt} but
vary from each other in treatment of nonperturbative hadronic contributions, see {\it e.g.} 
papers~\cite{Nomura:2012sb,Hagiwara:2011af} and references therein.
The Particle Data Group~\cite{Beringer:1900zz} quotes the following theoretical
estimate:
\begin{eqnarray}
a^{\mathrm{SM}}_\mu= 11\, 659\, 180.2(0.2)(4.2)(2.6)\cdot 10^{-10},
\end{eqnarray}
where the errors are given for the electroweak (EW) contribution, the lowest-order
hadronic one, and the higher-order hadronic one, respectively.
Let us note that the difference
\begin{eqnarray} \label{exp_SM}
\Delta a_\mu^{\mathrm{exp}-\mathrm{SM}} = 28.7(6.3)(4.9) \cdot 10^{-10}
\end{eqnarray}
is about twice as large as the weak (electroweak minus pure QED) contribution
\begin{eqnarray}
\Delta a_\mu^{\mathrm{weak}} = 15.4(0.2) \cdot 10^{-10}.
\end{eqnarray}
So, an explanation of the deviation by a contribution due to some
effects beyond the SM, requires introduction of a new energy scale 
being rather close to the EW one, while ongoing searches (in particular at LHC)
more and more disfavor finding new physics at such a scale. 
Attempts to explain the difference by some effect of strong interactions result 
in continuous efforts in calculations of the corresponding non-perturbative contributions 
and trying to fix them using relations to experimentally observables quantities, 
see {\it e.g.} review~\cite{Actis:2010gg}. Presently, there remains some valuable uncertainty 
in the QCD, but the possibility to explain in this way the difference~(\ref{exp_SM}) 
is very unlikely. Another possibility could be an error in the experimental analysis. 
It will be verified by the forthcoming experiments
Muon $g-2$~\cite{Roberts:2010cj} at Fermilab (USA) and J-PARC $g-2$~\cite{Mibe:2011zz} 
at KEK (Japan).   

In any case in order to resolve the puzzle, we have to look for the whole spectrum of possibilities. 
In this paper we suggest to discuss the scenario that some additional (but standard) interactions 
of muons within given experimental conditions could contribute to the observed gyromagnetic ratio. 
In particular, interactions of a muon with his neighbors in the beam can lead to a (small) effective 
shift of muons off their mass shells\footnote{Going off the mass shell due to emission of real (soft) 
photons (in vacuum) is forbidden by the energy-momentum conservation.}. 
Such collective effects due to electromagnetic interactions between charged particles in a beam have been
discussed in the literature, see {\it e.g.}~\cite{Ginzburg:1991ut}.
Remind also a weakened conservation of energy and momentum required to 
describe oscillations of atmospheric neutrinos originated from muon (and pion) decays. 

Let us state the following problem: how much do we need to shift a muon off its mass shell 
to get the observed shift in the anomalous magnetic moment?

\section{Off-shell muon form factor}

The one-loop electron from factor in the case of one external electron line
being off-shell (and the second one being on-shell) was considered in the 
book~\cite{Akhiezer:1959book}.
Here we will consider only the off-diagonal situation (one fermion off-shell and the other one on-shell). 
The change of the magnetic moment of a fermion which is permanently off-shell being in a 
central potential was treated in~\cite{Karshenboim:2005yf}. 

Let us consider one-loop QED corrections to the vertex of electron-photon interaction with the following 
kinematics:
\begin{eqnarray}
k^2 = 0,\qquad p^2 = - m^2, \qquad 
(p+k)^2+m^2 = 2pk = \kappa m^2,
\end{eqnarray}
where $k$ is the photon 4-momentum; $p$ and $p+k$ are
the electron momenta. One of them is off shell.
Note that metric $(-,+,+,+)$ is used. 
Dimensionless parameter $\kappa$ describes the degree
of the off-shellness, $\kappa<0$. We will consider small
values $|\kappa|\ll 1$. 

At the one-loop level, the fermion-photon vertex receives 
contributions from scalar, vector and tensor Feynman integrals 
$J_0$, $J_\sigma$, $J_{\sigma\tau}$. The scalar 
and the tensor integrals are infrared and ultraviolet divergent, respectively.
But as it is known from the standard calculations of $(g_f-2)$, just the vector integral
contributes to this quantity. For the off-mass-shell case it reads~\cite{Akhiezer:1959book} 
\begin{eqnarray}
\frac{m^2}{i\pi^2}J_\sigma = \biggl(J_0 - \frac{\ln|\kappa|}{\kappa-1}\biggr)p_\sigma
+ \biggl(2J_0 - \frac{\kappa-2}{\kappa-1}\ln|\kappa| - 2\biggr)\frac{k_\sigma}{\kappa},
\end{eqnarray}
where $J_0$ is the scalar integral,
\begin{eqnarray}
\frac{m^2}{i\pi^2}J_0 = \frac{1}{\kappa}\biggl[\Li2{1} - \Li2{1-\kappa}\biggr], \qquad
\Li2{x} \equiv -\int\limits_{0}^{x}\frac{\ln(1-y)}{y}dy.
\end{eqnarray}
In the limit $\kappa\to0$ and $k\to0$, the vector integral leads to the well
know one-loop result $\Delta a_f^{(1)} = \alpha/(2\pi)$ 
which was received first by J.~Schwinger~\cite{Schwinger:1948iu}.

Expanding in $\kappa$ we get the first correction due to the off-shellness:
\begin{eqnarray}
&& \Delta a_f^{(1,\kappa)} = \frac{\alpha}{2\pi}\biggl[1 + \delta a_f^{(\kappa)}\biggr], 
\nonumber \\
&& \delta a_f^{(\kappa)} = \biggl(\frac{1}{4}+\frac{\ln|\kappa|}{2}\biggr)\kappa 
+ \mathcal{O}(\kappa^2).
\end{eqnarray}

Assuming that the difference~\ref{exp_SM}) between experimental result and theoretical predictions
for the muon anomalous magnetic moment is due to the off-shellness effect, we get the equation 
to define the value of $\kappa$:
\begin{eqnarray}
\Delta a_\mu^{\mathrm{exp}-\mathrm{SM}}\approx 3 \cdot 10^{-9} = \frac{\alpha}{2\pi}\delta a_\mu^{(\kappa)}.
\end{eqnarray}
Numerical solution of the equation gives $\kappa\approx - 3.5\cdot 10^{-7}$.
Note that for small values of $\kappa$ the shift has the proper sign and the solution exists.
Such a value corresponds to off-shellness of a muon of the order $m|\kappa|\sim 35$~eV.

\section{Conclusion}

In this way we demonstrated the possibility to get a shift in the observed
value of the muon anomalous magnetic moment due an effective off-shellness of muons. 
To estimate the mean value of the latter one has to analyze concrete experimental 
conditions. Obviously, having the value of the beam density distribution
one would easily find the mean shift off the mass shell. 

The future experiment~\cite{Mibe:2011zz} utilizing an ultra-cold muon beam will have 
a completely different set-up in comparison to E821. 
Presumably, collective effects due to interactions of muons with each other in the beam will 
be very much suppressed in the Japanese experiment conditions.
The New Muon $g-2$ experiment at Fermilab~\cite{Roberts:2010cj} 
will have also somewhat different beam parameters with respect to the E821 one at BNL. 
So, the possible impact of collective interactions on $(g_\mu-2)$ will be accessed
in both experiments. 

Note also that the electron anomalous magnetic moment was measured in experiments on atomic spectroscopy 
where the off-shell effects are well under control.
 
\subsection*{Acknowledgments}
We are grateful to I.~Ginzburg, F.~Jegerlehner, S.~Karshenboim, and E.~Kuraev 
for fruitful stimulating discussions. One of us (A.A.) thanks for a financial support 
the Dynasty Foundation.


\begin{thebibliography}{99}

\bibitem{Beringer:1900zz}
  J.~Beringer {\it et al.}  [Particle Data Group Collaboration],
  Phys.\ Rev.\ D {\bf 86} (2012) 010001.

\bibitem{Aoyama:2012wj}
  T.~Aoyama, M.~Hayakawa, T.~Kinoshita and M.~Nio,
  Phys.\ Rev.\ Lett.\  {\bf 109} (2012) 111807
  [arXiv:1205.5368 [hep-ph]].

\bibitem{Bennett:2006fi}
  G.~W.~Bennett {\it et al.}  [Muon G-2 Collaboration],
  Phys.\ Rev.\ D {\bf 73} (2006) 072003
  [hep-ex/0602035].

\bibitem{Aoyama:2012wk}
  T.~Aoyama, M.~Hayakawa, T.~Kinoshita and M.~Nio,
  Phys.\ Rev.\ Lett.\  {\bf 109} (2012) 111808
  [arXiv:1205.5370 [hep-ph]].

\bibitem{Czarnecki:2002nt}
  A.~Czarnecki, W.~J.~Marciano and A.~Vainshtein,
  Phys.\ Rev.\ D {\bf 67} (2003) 073006
   [Erratum-ibid.\ D {\bf 73} (2006) 119901]
  [hep-ph/0212229].

\bibitem{Nomura:2012sb}
  D.~Nomura and T.~Teubner,
  Nucl.\ Phys.\ B {\bf 867} (2013) 236
  [arXiv:1208.4194 [hep-ph]].

\bibitem{Hagiwara:2011af}
  K.~Hagiwara, R.~Liao, A.~D.~Martin, D.~Nomura and T.~Teubner,
  J.\ Phys.\ G {\bf 38} (2011) 085003
  [arXiv:1105.3149 [hep-ph]].

\bibitem{Actis:2010gg}
  S.~Actis {\it et al.}  [Working Group on Radiative Corrections and Monte Carlo Generators for Low Energies Collaboration],
  Eur.\ Phys.\ J.\ C {\bf 66} (2010) 585
  [arXiv:0912.0749 [hep-ph]].

\bibitem{Roberts:2010cj}
  B.~L.~Roberts,
  Chin.\ Phys.\ C {\bf 34} (2010) 741
  [arXiv:1001.2898 [hep-ex]].

\bibitem{Mibe:2011zz}
  T.~Mibe [J-PARC g-2 Collaboration],
  Nucl.\ Phys.\ Proc.\ Suppl.\  {\bf 218} (2011) 242.
  
\bibitem{Ginzburg:1991ut}
  I.~F.~Ginzburg, G.~L.~Kotkin, S.~I.~Polityko and V.~G.~Serbo,
  Phys.\ Lett.\ B {\bf 286} (1992) 392
   [JETP Lett.\  {\bf 55} (1992) 637].

\bibitem{Akhiezer:1959book}
A.I. Akhiezer, V.B. Berestetskii, 
{\em Elements of Quantum Electrodynamics},
Oldbourne Press, 1964 [translated from the Russian edition (Moscow, ed. 2, 1959)].

\bibitem{Karshenboim:2005yf}
  S.~G.~Karshenboim, R.~N.~Lee and A.~I.~Milstein,
  Phys.\ Rev.\ A {\bf 72} (2005) 042101.

\bibitem{Schwinger:1948iu}
  J.~S.~Schwinger,
  Phys.\ Rev.\  {\bf 73} (1948) 416.


\end{thebibliography}
\end{document}